\documentclass[twocolumn, prl, showpacs]{revtex4}
\usepackage{amssymb}
\usepackage{graphicx}
\usepackage{dcolumn}
\usepackage{bm}
\usepackage{amsmath}
\usepackage[colorlinks,linkcolor=red,citecolor=blue]{hyperref}

\begin{document}

\title{Spin exchange-induced spin-orbit coupling in a superfluid mixture}
\author{Li Chen$^{1,2}$}
\author{Chuanzhou Zhu$^{1}$}
\author{Yunbo Zhang$^{2}$}
\email{ybzhang@sxu.edu.cn}
\author{Han Pu$^{1,3}$}
\email{hpu@rice.edu}

\affiliation{
$^1${Department of Physics and Astronomy, and Rice Center for Quantum
    Materials, Rice University, Houston, TX 77005, USA}\\
$^2${Institute of Theoretical Physics, Shanxi University, Taiyuan, Shanxi
030006, P. R. China}\\
$^3${Center for Cold Atom Physics, Chinese Academy of Sciences, Wuhan
430071, China}}

\begin{abstract}
We investigate the ground-state properties of a dual-species spin-1/2 Bose-Einstein condensate. One of the species is subjected to a pair of Raman laser beams that induces spin-orbit (SO) coupling, whereas the other species is not coupled to the Raman laser. In certain limits, analytical results can be obtained. It is clearly shown that, through the inter-species spin-exchange interaction, the second species also exhibits SO coupling. This mixture system displays a very rich phase diagram, with many of the phases not present in an SO coupled single-species condensate. Our work provides a new way of creating SO coupling in atomic quantum gases, and opens up a new avenue of research in SO coupled superfluid mixtures. From a practical point of view, the spin exchange-induced SO coupling may overcome the heating issue for certain atomic species when subjected to the Raman beams. 
\end{abstract}

\pacs{03.75.Mn, 67.60.Bc, 67.85.Fg}
\maketitle

\textit{Introduction }--- Since the first experimental realization of the synthetic spin-orbit (SO) coupling induced by a pair of Raman beams in a Rb Bose-Einstein condensate (BEC) \cite{Lin2011}, SO coupled quantum gases has emerged as one of the most active frontiers in cold atom research. SO coupling has been realized in both bosonic \cite{Lin2011} and fermionic \cite{Cheuk2012,Wang2012} atoms. Various theoretical models based on the Raman process have been proposed \cite{Campbell2011} and are predicted to give rise to a variety of rich many-body quantum phases \cite{Galitski2013, Goldman2014, Zhai2015}. One big challenge in practice is to overcome the heating problem induced by the Raman laser beams \cite{Wei2013, Zhai2015}. For certain atomic species, this heating can be very severe such that prevents the realization of many interesting quantum phases.

Another research frontier in cold atoms is superfluid mixtures, a topic that has been studied in the context of superfluid $^3$He-$^4$He mixtures in condensed matter physics for many decades \cite{Guenault1983, Tuoriniemi2002}. Dual-species atomic BECs have been realized quite a few years ago \cite{Hall1998, Maddaloni2000}. More recently, several groups have successfully realized superfluid mixtures consisting of one bosonic and one fermionic species \cite{Barbut2014, Yao2016, Roy2017}. The main motivation for such studies is to investigate the new phases induced by inter-species interaction that are not present in single-species systems.  

In this work, we consider a two-species spinor BEC \cite{Ho1996, Pu1998, Law1998, Xu2009, YShi2010} and investigate the effects of SO coupling in such a system. More specifically, each species in our study represents a spin-1/2 condensate, as produced in a recent experiment \cite{XLi2015}. One of the species is subjected to a Raman-induced SO coupling \cite{Lin2011, Ho2011, YLi2012, Zheng2013}, whereas the other species is not directly coupled to the Raman beams. We show that the latter species acquires an effective SO coupling due to the spin-exchange interaction between the two. From a practical point of view, such spin exchange-induced SO coupling may overcome the heating problem suffered by certain atomic species when subjected to Raman beams. From a more fundamental point of view, SO coupling in spinor superfluid mixtures represents a new area of research in this field.

\begin{figure}[t]
\includegraphics[width=0.48\textwidth]{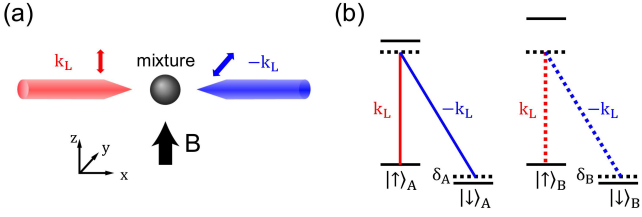}
\caption{(a) Schematic representation of the dual-species BEC system subjected to a pair of Raman beams. (b) Atomic level structure for either species. We assume in this work that the Raman coupling strength $\Omega_B$ for species $B$ is negligible \cite{RamanB}.}
\label{FigSetup}
\end{figure}

\textit{Model }--- We consider a dual-species condensate mixture, labeled as $A$ and $B$, with each species representing a spin-1/2 system with two internal states labeled as $\uparrow$ and $\downarrow$. Species $A$ is subjected to a pair of laser beams which induces a Raman transition between its two spin states and imposes a momentum recoil $\pm\hbar k_L$ along the $x$-axis, as schematically shown in Fig.~\ref{FigSetup}(a). For simplicity, we assume the condensates to be tightly confined along the other two directions, and can be considered as quasi-one dimensional. In the mean-field framework, the total energy functional of the system reads (we set $\hbar=1$):
\begin{equation}
  \label{eq:H}
 E\left(\mathbf{\Psi}_A,\mathbf{\Psi}_B\right) = \sum_{i=A,B}{\int{dx}\mathbf{\Psi}_i^\dagger h_i \mathbf{\Psi}_i}+ \sum_{i=A,B}{\mathcal{G}_i}+\mathcal{G}_{AB},
\end{equation}
 where $\mathbf{\Psi}_i=\left( \psi _{i,\uparrow },\psi _{i,\downarrow }\right)^{T}$ represents the spinor wave function satisfying $\int{dx\left|\mathbf{\Psi}_i\right|^2=N_i}$ with $N_i$ being the total particle number of species $i$ ($i=A$ or $B$), and
\begin{equation}
   \label{eq:h}
   h_i=\frac{k_{i}^2}{2m_i}+\frac{\delta_i}{2}\sigma_{z}^{i}+\frac{\Omega_i}{2}
   \begin{pmatrix}
     0 & e^{-2ik_Lx} \\
     e^{2ik_Lx} & 0 \\
   \end{pmatrix}
   +V^i_{\rm ext}\left(x\right)
\end{equation}
is the single-particle Hamiltonian where $\delta_i$ denotes the two-photon detuning, $V^i_{\rm ext}\left(x\right)$ is the external potential, and $\Omega_i$ characterizes the two-photon Raman coupling strength. In this work, we will take $\Omega_B=0$, i.e., the Raman beams do not interact directly with species $B$ \cite{RamanB}. Furthermore, we will focus on the case with $\delta_A=\delta_B=0$, which corresponds to a situation where the bare atomic energy difference between the two spin states are the same for both species, and the Raman transition in species $A$ is on resonance \cite{note}.  

The last two terms in Eq.~(\ref{eq:H}) characterize two different types of two-body interactions. The intra-species interaction is given by 
\begin{equation}
\mathcal{G}_i=\frac{1}{2}\int{dx\left[g^i\left(\rho_{i,\uparrow}^2+\rho_{i,\downarrow}^2\right)+2g^i_{\uparrow\downarrow}\rho_{i,\uparrow}\rho_{i,\downarrow}\right]}\,,
\end{equation}
where $\rho_{i,\sigma} = \left|\psi_{i,\sigma}\right|^2$ is the density of spin-$\sigma$ for species $i$, and we have assumed that the interaction between like spins has strength $g^i$ independent of the spin for both species. The inter-species interaction takes the form
\begin{equation}
\mathcal{G}_{AB}=\int dx\left[ \gamma \rho_{A}\rho_{B}+\beta\left(\psi_{A,\uparrow}^{*}\psi_{B,\downarrow}^{*}\psi_{B,\uparrow}\psi_{A,\downarrow}+c.c. \right)\right] \,,\label{eq:GAB}
\end{equation}
where the density-density interaction term, characterized by strength $\gamma$, is assumed to be spin-independent, with $\rho_i= \rho_{i,\uparrow}+\rho_{i,\downarrow}$ the total density for species $i$. The term characterized by $\beta$ is the spin-exchange term which, as we will show, plays a crucial role in our study. 

To make the system stable, we assume that all density-density interactions are repulsive with positive interaction strengths (i.e., $g^i, \, g^i_{\uparrow \downarrow}, \,\gamma >0$). Furthermore, we want all the components to be miscible, hence take $g^i_{\uparrow \downarrow} < g^i$ and $\gamma < \sqrt{g^A g^B}$ \cite{Pethick2008}. The spin-exchange interaction strength $\beta$ can, in principle, be complex. However, its phase angle can be taken to be zero by a simple redefinition of the atomic wave functions. As a result, we will assume $\beta >0$ without loss of generality. 

Following a standard procedure, we take the gauge transformation for the wave function as $\mathbf{\Psi}_i \longrightarrow U_i \mathbf{\Psi}_i$ with $U_i = e^{ik_Lx\sigma_z^i}$. Under this transformation, the interaction terms are invariant, and the single-particle Hamiltonians take the new form as (with $\Omega_B=0$ and $\delta_i=0$):
\begin{eqnarray}
h_A &=& \frac{\left(k_{A}-k_{L}\sigma_z^A \right)^2}{2m_A}+\frac{\Omega_A}{2}\sigma_{x}^A+V^A_{\rm ext}\left(x\right) \,,\label{eq:hA}\\
h_B &=& \frac{\left(k_{B}-k_{L}\sigma_z^B\right)^2}{2m_B}+V^B_{\rm ext}\left(x\right) \,.\label{eq:hB}
\end{eqnarray}
With a finite $\Omega_A$, $h_A$ contains an effective non-Abelian gauge field leading to the SO coupling. Whereas, no SO coupling is present in $h_B$ due to the lack of the Raman coupling. However, as we will show below, the spin-exchange term included in $\mathcal{G}_{AB}$ will provide an effective Raman coupling, and hence induce an effective SO coupling, for species $B$.

\textit{Case of }$N_A\gg N_B$ --- To clearly demonstrate how the spin-exchange interaction induces SO coupling in species $B$, let us first consider a situation with $N_A \gg N_B$ such that the effect of $B$ on $A$ can be neglected to a good approximation. As a result, the properties of $A$ are the same as in a single-species SO coupled condensate, which has been extensively studied in previous works. Furthermore, we will first assume that $V_{\rm ext}^{A,B}=0$ which allows us to find analytical solutions. For such a homogeneous quasi-one dimensional system, $A$ possesses three mean-field phases \cite{YLi2012} separated by two critical values of Raman coupling strength $\Omega _{A}^{S-P}$ and $\Omega _{A}^{P-Z}$ that are give by
\begin{eqnarray}
\Omega_{A}^{S-P}\!\! &=& \!\! 2\left[\left(k_L^2+G^A_+\right)\left(k_L^2-2G^A_-\right)\frac{2G^A_-}{G^A_++2G^A_-}\right]^{1/2},\label{eq:OmgCSP} \\
\Omega_{A}^{P-Z} \!\! &=& \!\! 2\left(k_L^2-2G^A_-\right)\,,\label{eq:OmgCPZ}
\end{eqnarray}
where $G^A_{\pm} \equiv \rho_0\left(g^A \pm g^A_{\uparrow\downarrow}\right)/4$, with $\rho_0$ being the average total density.

For small coupling strength $\Omega_A < \Omega _{A}^{S-P}$, $A$ is in the stripe phase (ST) whose wave function can be approximately depicted by a superposition of two plane waves with all higher-order harmonic terms neglected \cite{YLi2013, Yamamoto2017}      
\begin{equation}
\label{eq:PsiA}
\mathbf{\Psi}_{A}^{\rm ST}= \sqrt{\frac{\rho_0}{2}}\left[
\begin{pmatrix}
\cos \theta_A \\
-\sin \theta_A%
\end{pmatrix}%
e^{i\kappa_Ax}+
\begin{pmatrix}
\sin \theta_A \\
-\cos \theta_A%
\end{pmatrix}%
e^{-i\kappa_Ax}\right],
\end{equation}
where $\kappa_A$ is finite and $2\theta_A=\cos^{-1}\left(\kappa_A/k_L\right)$, and the corresponding density profiles oscillate in space and are given by
\begin{equation}
\rho_{A,\uparrow}(x) = \rho_{A, \downarrow} (x)=\frac{\rho_A(x)}{2}= \frac{\rho_0}{2} [1+ \sin 2\theta_A \cos(2\kappa_A x)] \,. \label{rhost}
\end{equation}
For intermediate coupling strength $\Omega_{A}^{S-P}<\Omega_A <\Omega_{A}^{P-Z}$, $A$ is in the plane-wave phase (PW) whose ground state is doubly degenerate with 
\begin{equation}
\mathbf{\Psi}_{A}^{\rm PW} \!= \!\sqrt{\rho_0}
\begin{pmatrix}
\cos \theta_A \\
-\sin \theta_A%
\end{pmatrix}%
\! e^{i\kappa_Ax} \,, \;{\rm or}\;
\sqrt{\rho_0} \begin{pmatrix}
\sin \theta_A \\
-\cos \theta_A%
\end{pmatrix}%
\! e^{-i\kappa_Ax}.  \label{eq:PW}
\end{equation}
Finally, for large coupling strength $\Omega_A >\Omega_{A}^{P-Z}$, $A$ is in the zero-mode phase (ZM) with $\mathbf{\Psi}_{A}^{\rm ZM}= \sqrt{\rho_0/2}\, (1,-1)^T$, which has the same form as the wave function in PW phase with $\kappa_A=0$ and $\theta_A=\pi/4$. In both the PW and the ZM phases, species $A$ possesses homogeneous density profiles.

\begin{figure}[ht]
	\includegraphics[width=0.45\textwidth]{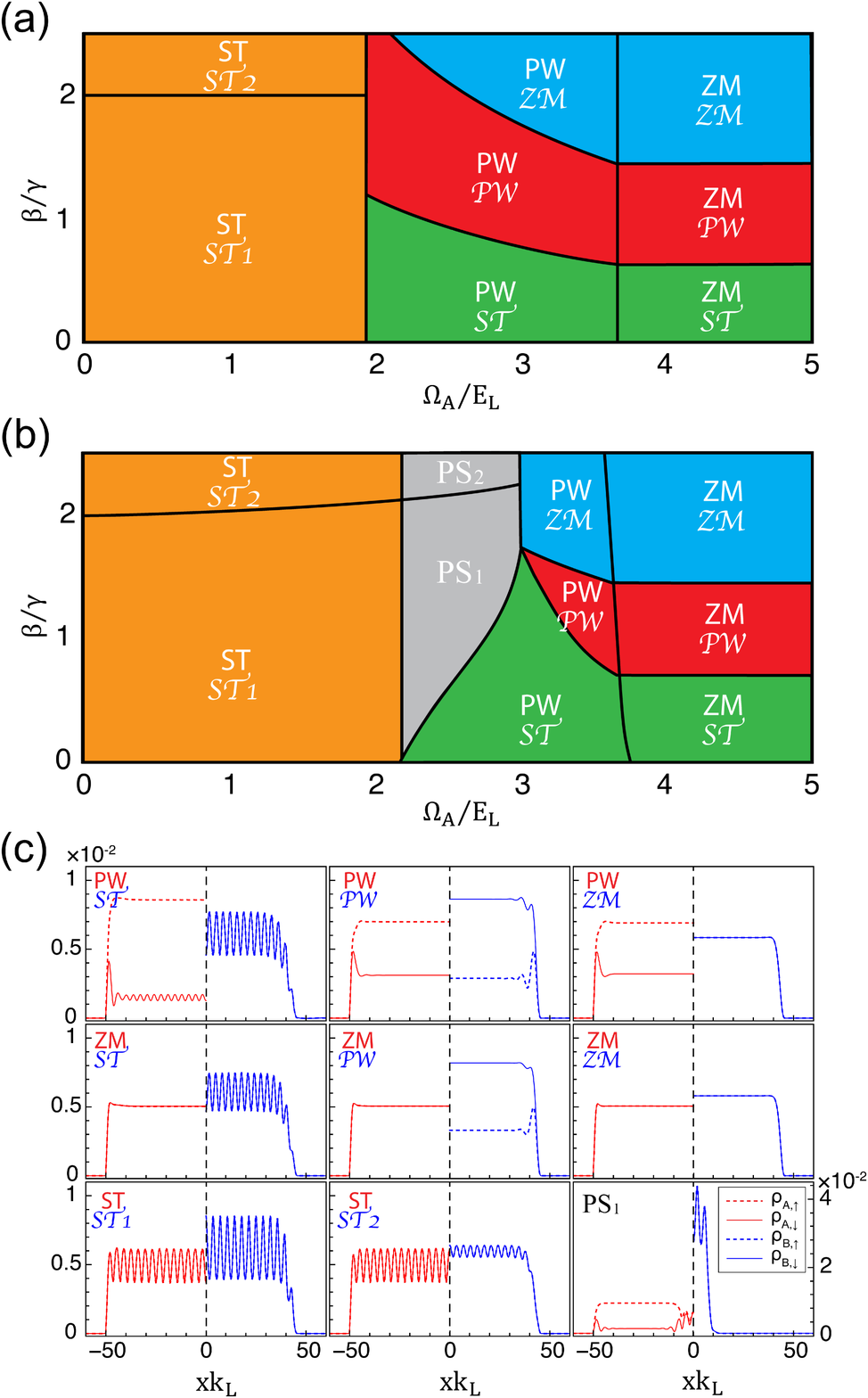}
	\caption{(a), (b) Ground-state phase diagram obtained analytically and numerically, respectively, in the case of $N_A\gg N_B$. (c) Typical density profiles for various phases. $\Omega_A$ is in units of the photon recoil energy $E_L=k_L^2/2m_A$. In each subplot in (c), we plot only the left (right) half of the density profiles for species $A$ ($B$), as the full profile is mirror symmetric about $x=0$. In this calculation, we take $m_A$ and $m_B$ to be the mass of $^{23}$Na and $^{87}$Rb, respectively; and $N_A=2.5\times10^4=25 N_B$. The box potential has a length of $L_A=100k_L^{-1}\approx 27\mu m$ for species $A$. To minimize the edge effects, we choose the box length for $B$ to be slightly smaller with $L_B=0.9L_A$. The interaction strengths are: $g^A=6.48\times10^{-3}E_{L}/k_{L}$, $g^B=3.44\times10^{-3}E_{L}/k_{L}$, $g^i_{\uparrow\downarrow}=0.8g^i$, $\gamma = 2.87\times10^{-3}E_{L}/k_{L}$. The densities in (c) are in units of $k_{L}$, and are renormalized such that $\int dx \rho_i(x) =1$. }
	\label{FigPD1}
\end{figure}

Now let us turn our focus onto species $B$, which is influenced by $A$ through the inter-species interaction term  $\mathcal{G}_{AB}$ in Eq.~(\ref{eq:GAB}). First, consider $A$ to be in one of the homogeneous phases (PW or ZM), in which its total density is constant $\rho_A=\rho_0$. Inserting $\mathbf{\Psi}_{A}$ into (\ref{eq:GAB}), we arrive at an effective single-particle Hamiltonian for species $B$ as: 
\begin{equation}
  \label{eq:hBeff}
   h_{B,{\rm eff}}=\frac{\left(k_{B}-k_{L}\sigma_z^B\right)^2}{2m_B}+\frac{\Omega_B}{2}\sigma_{x}^B,
 \end{equation}
where 
\begin{equation}
  \label{eq:OmegaB}
  \Omega_B = -\beta \rho_0 \, \sin{2\theta_A}.
\end{equation}
It is clear from Eq.~(\ref{eq:hBeff}) that the spin-exchange interaction provides an effective Raman coupling with strength $\Omega_B$, and as a consequence, species $B$ also experiences SO coupling as the form of $h_{B,{\rm eff}}$ is identical to that of $h_A$. We expect that $B$ too possesses the three phases $\mathcal{ST}$, $\mathcal{PW}$ and $\mathcal{ZM}$, depending on the magnitude of $\beta$ which determines $\Omega_B$. In analogy to Eqs.~(\ref{eq:OmgCSP}) and (\ref{eq:OmgCPZ}), we can find the two critical values of $\beta$ that separates the three phases for Species $B$ as
\begin{eqnarray}
\label{eq:OmgCSPB}
\beta_{B}^{S-P}\!\! &=& \!
\frac{2}{\rho_0 \sin{2\theta_A}} \left[
\frac{2G^B_-\left(k_L^2+G^B_+\right)\left(k_L^2-2G^B_-\right)}{\left(G^B_++2G^B_-\right)}\right]^{1/2}\,,\\
\label{eq:OmgCPZB}
\beta_{B}^{P-Z} \!\! &=& \! \frac{2\left(k_L^2-2G^B_-\right)}{\rho_0 \sin{2\theta_A}}\,, 
\end{eqnarray}
where $G^B_{\pm}=\rho_1\left(g^B \pm g^B_{\uparrow\downarrow}\right)/4$, with $\rho_1$ being the average total density of species $B$.
In each of these phases, the wave function $\mathbf{\Psi}_{B}$ should have a similar form to $\mathbf{\Psi}_{A}$ in the corresponding phase.

When species $A$ is in the ST phase, following a similar procedure as above, we can obtain the effective Hamiltonian for $B$ as
\begin{equation}
h'_{B,{\rm eff}}=\frac{\left(k_{B}-k_{L}\sigma_z^B\right)^2}{2m_B}+\rho_A(x)\,\begin{pmatrix}
\gamma & -\frac{\beta e^{i\phi(x)}}{2}\\
-\frac{\beta e^{-i\phi(x)}}{2} & \gamma
\end{pmatrix} , \nonumber
\end{equation}
where $\rho_A(x)$ is the density profile of $A$ given in Eq.~(\ref{rhost}), and
\begin{equation}
\label{eq:phi}
\phi(x) = \text{tan}^{-1} {\left(\frac{ \cos{2\theta_A} \, \sin{2\kappa_A x}} {\sin{2\theta_A}+\cos{2\kappa_A x}}\right)}.
\end{equation}
In general, the sinusoidal oscillation in $\rho_A$ also leads to a sinusoidal oscillation in $B$ and drives the latter to a stripe phase. Whether the density stripes in $A$ and in $B$ are in phase or not can be roughly determined as follows. Diagonalizing the matrix in the second term of $h'_{B,{\rm eff}}$, we readily find its lower eigenvalue as $\gamma -\beta/2$. Hence we can regard $(\gamma-\beta/2) \rho_A(x)$ as an effective potential for atoms in species $B$. As a result, the density stripes in the two species will be in phase if $\gamma-\beta/2<0$, and out of phase otherwise. 

Through these considerations, we obtain analytically the ground state phase diagram in the $\Omega_A - \beta$ parameter space as shown in Fig.~\ref{FigPD1}(a). Furthermore, to confirm the analytical results, 
we directly solve the coupled Gross-Pitaevskii equation derived from the energy functional in Eq.~(\ref{eq:H}). In the numerical calculation, we include a box potential with hard walls for both species \cite{notebox}. We present the numerically obtained phase diagram in Fig.~\ref{FigPD1}(b), with typical density profiles for different phases plotted in Fig.~\ref{FigPD1}(c). In the phase diagram, each phase of the mixture system, bounded by solid black lines, is labeled by the corresponding phase of individual species. For example, ZM/$\mathcal{ST}$ is the phase where species $A$ is in the ZM phase and $B$ in the $\mathcal{ST}$ phase (except for the two phases labeled PS$_1$ and PS$_2$ in Fig.~\ref{FigPD1}(b), see below). At large $\Omega_A$, species $A$ is in the ZM phase, as $\beta$ increases, the effective Raman coupling strength $\left|\Omega_B\right|$ [see Eq.~(\ref{eq:OmegaB})] increases, and species $B$ goes through $\mathcal{ST}$, $\mathcal{PW}$ and $\mathcal{ZM}$ phases (see the right part of the phase diagram), the corresponding density profiles are plotted in the middle row of Fig.~\ref{FigPD1}(c). At intermediate $\Omega_A$, species $A$ is in the PW phase, and as $\beta$ increases, species $B$ again goes through $\mathcal{ST}$, $\mathcal{PW}$ and $\mathcal{ZM}$ phases (see the middle part of the phase diagram), the corresponding density profiles are plotted in the upper row of Fig.~\ref{FigPD1}(c). For small $\Omega_A$, species $A$ is in the ST phase, and species $B$ also exhibits density stripe regardless of the value of $\beta$. For small $\beta$, the stripes of the two species are out of phase, and we label the phase of the combined system as ST/$\mathcal{ST}_1$. For large $\beta$, the stripes of the two species are in phase, and we label the phase of the combined system as ST/$\mathcal{ST}_2$. The boundary between ST/$\mathcal{ST}_1$ and ST/$\mathcal{ST}_2$ is given by $\beta/\gamma =2$ in the analytical calculation. The numerical calculation shows that this boundary has a weak dependence on $\Omega_A$. 

Comparing Fig.~\ref{FigPD1}(a) and (b), one can see that, the two phase diagrams are in general in good qualitative agreement. The main difference is that the analytical phase diagram does not produce the two phases labeled as PS$_1$ and PS$_2$ in the middle of the numerical phase diagram.
These two phases correspond to phase separation. In PS$_1$, species $B$ mainly occupy the middle of the box potential and, together with species $A$, forms the ST/$\mathcal{ST}_1$ phase. At the edges of the box, we have only species $A$ which is in the PW phase. Typical density profiles of PS$_1$ are presented in the lower right sub-plot of Fig.~\ref{FigPD1}(c). PS$_2$ is similar to PS$_1$, only that we have the ST/$\mathcal{ST}_2$ phase in the middle of the box. That these two phases are not present in Fig.~\ref{FigPD1}(a) can be attributed to the breakdown of the assumption that species $A$ is not affected by $B$, which underlies the analytical calculation. This assumption generally holds when species $A$ is in either ST or ZM modes. However, when species $A$ is in the PW phase, it can possess a large spin polarization. Hence even though the total atom number $N_A$ is much larger than $N_B$, the number in the minority spin component of $A$ may be comparable to $N_B$. As a result, species $B$ may have a significant influence on $A$. 

\begin{figure}[t]
	\includegraphics[width=0.45\textwidth]{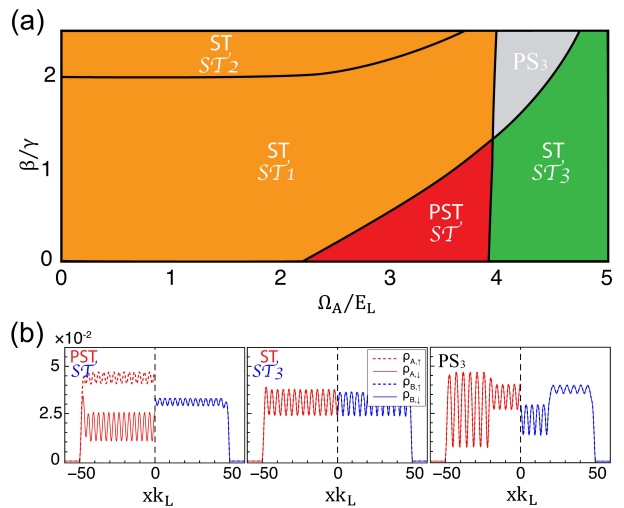}
	\caption{(a) Ground-state phase diagram, and (b) typical density profiles for various phases, in the case of $N_A= N_B=1\times10^4$. Both species are confined in a box potential with length $L_A=L_B=100k_L^{-1}$. Other parameters used are the same as in Fig.~\ref{FigPD1}. Note that the PST/$\mathcal{ST}$ phase has two degenerate ground states. The one not shown has the same density profiles in species $B$, but reversed density distributions for $\rho_{A,\uparrow}$ and $\rho_{A,\downarrow}$ as plotted in the first panel of (b).}
	\label{FigPD2}
\end{figure}

\textit{Case of }$N_A = N_B$ --- The situation discussed above provides a clear picture how the spin-exchange interaction can induce SO coupling in species $B$. Now let us consider the situation where $N_A=N_B$. Now the mutual influence between the two species is important, and we have to resort to numerical calculations to investigate this system. The phase diagram and several representative density profiles are presented in Fig.~\ref{FigPD2}. In the calculation, all the parameters are kept the same as in Fig.~\ref{FigPD1}, only that $N_B$ is increased to be equal to $N_A$. One striking feature one can immediately notice from the phase diagram is that the stripe phase dominates the parameter space. The left region of the phase diagram (small $\Omega_A$) is still occupied by ST/$\mathcal{ST}_1$ and ST/$\mathcal{ST}_2$ as in the previous case, but the regions for both phases are much enlarged. In a single species Raman-induced SO coupled condensate, the ST phase occurs at small Raman coupling strength. The enlarged stripe phase region in the mixture may be attributed to the fact that the back action from species $B$ reduces the effective Raman coupling in species $A$. 

The other three phases, PST/$\mathcal{ST}$, ST/$\mathcal{ST}_3$, and PS$_3$, in Fig.~\ref{FigPD2}(a) do not exist in the previous case. In PST/$\mathcal{ST}$, species $B$ is a conventional stripe phase whose wave function is approximately given by Eq.~(\ref{eq:PsiA}), representing an equal-weight superposition of two plane-wave states with opposite momenta. By contrast, the state of species $A$ with two-fold degeneracy can be roughly regarded as an unequal-weight superposition of two plane wave states as given in Eq. (\ref{eq:PW}). In other words, $A$ is roughly a hybrid of the ST and the PW state. The density profiles of $A$ exhibit stripes but one spin state has more population than the other, as shown in the left subplot in Fig.~\ref{FigPD2}(b).  

In ST/$\mathcal{ST}_3$ which occurs at large $\Omega_A$, species $B$ is still roughly a conventional stripe phase, but the wave function of species $A$ takes the approximate form as follows:
\begin{equation}
\label{eq:PsiA2}
\mathbf{\Psi}_{A}\propto C_1
\begin{pmatrix}
1 \\
-1%
\end{pmatrix}%
+C_2
\begin{pmatrix}
e^{2i\kappa_Ax}\\
-e^{-2i\kappa_Ax}
\end{pmatrix}\,,
\end{equation}
which can be regarded as a hybrid of the ZM and the PW phase. 
Finally, PS$_3$ is a phase separated state, where ST/$\mathcal{ST}_3$ occupies the middle of the box potential, and ST/$\mathcal{ST}_1$ occupies the edges.

\textit{Conclusion and Outlook} --- To summarize, we have presented a study of a mixture of two spin-1/2 condensates, with only one of the species subjected to a pair of Raman laser beams which induces SO coupling in that species. Through the inter-species spin-exchange interaction, however, the other species also exhibits SO coupling. With many control parameters, such as the relative atomic numbers, interaction strengths, etc., the mixture system displays a very rich phase diagram, and many of the phases do not exist in a single-species system. 

From a practical point of view, our method provides a viable way of achieving SO coupling in species that suffer from severe Raman-induced heating \cite{Wei2013, Zhai2015}. From a fundamental point of view, our work opens up a new avenue of research in the study of SO coupling in atomic quantum gases. We have considered here, perhaps, the simplest spinor mixtures. This can be naturally extended to mixtures of high spin systems \cite{Campbell2016, Lan2014, Natu2015, Martone2016, Sun2016, Yu2016, Chen2016}, where both the inter- and intra-species spin-exchange interactions exist, the interplay between which may lead to even richer physics. The system considered here is quasi-one dimensional. Extending our calculation to higher dimensions \cite{Wang2010, Wu2011, Hu2012, Chen2015, Li2016} may lead to the realization of new types effective SO coupling. Finally, similar study can also be extended to Bose-Fermi mixtures \cite{Barbut2014, Yao2016, Roy2017}, which will be extremely important as the most commonly used fermionic species, such as $^6$Li \cite{Cheuk2012}and $^{40}$K \cite{Wang2012}, all suffer significant Raman-induced heating, and we are still waiting for the first experimental achievement of SO coupled superfluid Fermi gas.

\begin{acknowledgments}
L. Chen would like thank L. Dong for helpful discussion of the computational codes. LC acknowledges support from the program of China Scholarship Council (No. 201608140089). HP acknowledges support from US NSF and the Welch Foundation
(Grant No. C-1669). YZ is supported by NSF of China under Grant Nos. 11234008, 11474189 and 11674201.
\end{acknowledgments}


\begin{thebibliography}{99}
\bibitem{RamanB} The two-photon Raman coupling strength depends on single-photon detuning and single-photon Rabi frequencies between the two electronic ground states and excited states, which are all species-dependent. It is possible that the same pair of laser beams induce a significant Raman transition in one species, but not in the other. In this work, we assume that the Raman coupling strength $\Omega_B$ for species $B$ can be neglected.
\bibitem{note} In practice, this can be achieved by properly tuning the Raman beam frequency, and by adding an external magnetic field to tune the atomic energy levels via the Zeeman effect. The essential physics we want to explore would not be qualitatively affected when the condition $\delta_i=0$ is relaxed a little.
\bibitem{notebox}We use the box potential so that the numerical results can be better compared with the analytical results obtained for a homogeneous system. We also carried out numerical calculations with harmonic potentials, and made sure that the physics is not qualitatively affected.
\end{thebibliography}
\end{document}